\documentstyle[epsf,sprocl]{article}

\def\Journal#1#2#3#4{{#1} {\bf #2}, #3 (#4)}

\def\be{\begin{equation}}
\def\ee{\end{equation}}
\def\bea{\begin{eqnarray}}
\def\eea{\end{eqnarray}}
\begin{document}

\title{Solitons in SO(5) Superconductivity}

\author{R. MacKENZIE}

\address{Laboratoire Ren\'e-J.-A.-L\'evesque, Universit\'e de
Montr\'eal,\\ Montr\'eal, Qu\'ebec H3C 3J7\\and\\
Department of Physics and Astronomy, University of British
Columbia,\\6224 Agriculture Rd, Vancouver BC V6T 1Z4}

\author{J.M. CLINE}
\address{Department of Physics, McGill University,\\
3600 University St., Montr\'eal, Qu\'ebec H3A 2T8}


\maketitle\abstracts{ 
A model unifying superconductivity and antiferromagnetism using an
underlying approximate SO(5) symmetry has injected energy into the
field of high-temperature superconductivity. This model might lead to
a variety of interesting solitons. In this paper, the idea that
superconducting vortices may have antiferromagnetic cores is
presented, along with the results of some preliminary
numerical work. An outlook for future work, including
speculations
about other possible exotic solitons, is presented.}

\section{Introduction}

The phase diagrams of a variety of exotic superconductors
(high-temperature superconductors, heavy fermion superconductors,
organic superconductors) have a very rich structure. Although profound
differences in these phase diagrams exist, it is surprising that in
all of them two features are common: superconductivity and
antiferromagnetism. It is extremely enticing to speculate that these
materials, despite the incredible range of underlying structures, may
have some common underlying reason for the appearance of these two
phases.

This idea was formalized by S.C. Zhang,\cite{zhang} who observed that
both superconductivity and antiferromagnetism involve spontaneous
symmetry breaking. Superconductivity is essentially spontaneous
breaking of electromagnetism (it is, in fact, the first example of
what we now call dynamical symmetry breaking); antiferromagnetism is
spontaneous breaking of spin-rotation symmetry. The first symmetry
group is U(1), or equivalently SO(2); the second is SO(3). Zhang's
suggestion, borrowing heavily on ideas from particle physics,
was that these two symmetries might be unified into a larger symmetry
group. He presented a strong case for the group SO(5). His work has
given rise to a minor cottage industry of SO(5) phenomenology, not to
mention fueling a heated debate (with some of the heavyweights of
condensed matter physics appearing on opposite sides) over the merits
and possible fundamental flaws of the idea.

In this work (which, admittedly, uses a rather broad interpretation of
the theme of this Institute -- ``Electroweak Physics''), the ABCs of
superconductivity and antiferromagnetism will be briefly reviewed, and
Zhang's unified description of the two will be outlined. The
case for exotic solitons will then be discussed. Superconducting
vortices with antiferromagnetic cores will be examined in some detail
(though much work remains), and other, even more speculative,
possibilities will be discussed.

\section{The SO(5) Model}

\subsection{Superconductivity}
Superconductivity is a phenomenon which occurs in a huge number of
materials, if a sufficiently low temperature is
reached. Superconductors display several striking features, among them
zero resistance, the Meissner effect, the existence of a gap in the
spectrum of low-energy excitations (this is not always
the case, but usually it is
so), and a transition to a normal state as the
temperature increases.

Most of these phenomena can be described by a phenomenological model,
known as a Ginzburg-Landau model. The great success of many workers in
the late fifties and early sixties, culminating in the work of
Bardeen, Cooper and Schrieffer (who first succeeded in an
essentially complete description of all ``conventional''
superconductors), was the derivation of this GL model from an
underlying microscopic model.

For our purposes, the GL model of superconductivity is essentially a
nonrelativistic version of what we in particle physics call the
Abelian Higgs model. The distinction between nonrelativistic and
relativistic models is irrelevant here, so I will discuss only the
relativistic case. The Lagrangian is
\be
{\cal L}=|D_\mu\phi|^2-{\lambda\over 4}\left(|\phi|^2-v^2\right)^2
-{1\over4}{F_{\mu\nu}}^2,
\label{gl}
\ee
where $D_\mu\phi=\partial_\mu\phi-i2eA_\mu\phi$ and where
$\phi=c_{p,\uparrow}^\dagger c_{-p,\downarrow}^\dagger$ is the Cooper
pair creation operator.

When $\langle\phi\rangle\neq0$ (as is the case in the above
Lagrangian), the O(2) symmetry is spontaneously broken, and it is easy
to derive such features as the Meissner effect and a gap in the
excitation spectrum from this fact.

\subsection{Antiferromagnetism}

In elementary discussions of spontaneous symmetry breaking, often the
first example given is the ferromagnet. At high temperatures a spin
system has spins oriented randomly. As the temperature is decreased,
it can occur that the spins prefer to be aligned in the same direction
as their neighbours. This alignment amounts to the selection of an
{\it a
priori} random
direction in space which is singled out. Rotations of the resulting
configuration about this direction do not alter the system (that
symmetry is not broken), yet rotations about any other direction do
change the system (those symmetries are broken). Thus the ferromagnet
breaks spin rotational symmetry from SO(3) (rotations about any
direction when the average magnetism in any region is zero) to SO(2)
(rotations about the direction in which the spins are aligned).

Somewhat less familiar, but no more complicated (for our purposes, at
least),
is the case of antiferromagnetism. There, adjacent spins prefer to be
anti-aligned at low energies. Once one spin's direction is chosen, all
the others must follow suit (alternating in direction from site to
site) in order to minimize the energy. Once again, SO(3) spin rotation
symmetry is broken to SO(2).

The order parameter in the case of ferromagnetism is (somewhat
loosely) the average value of the spin $\langle \vec S\rangle$; in the
case of antiferromagnets this averages to zero, but one defines a
staggered spin vector $\vec n=\langle (-)^n\vec S\rangle$, where the extra
sign is positive or negative depending whether one is on a site an
even or odd number of translations away from some reference site.

\subsection{Combined superconductivity and antiferromagnetism}

We have seen that superconductivity can be described by a complex
field $\phi$ (or, equivalently, by a real doublet of fields defined by
$\phi=(\phi_1+i\phi_2)/\sqrt{2}$), and that antiferromagnetism can be
described by a real triplet field $\vec n$. Zhang put forth the idea
that, since these two order parameters seem to be relevant to such a
wide variety of systems, perhaps there is an underlying (approximate)
symmetry which includes these two as subgroups, rather like the
central idea
of Grand Unified Theories. This theory would then be described by a
five-component vector $\vec N=(n_1,n_2,n_3,\phi_1,\phi_2)$; if the
dynamics of the system is such that the expectation value of $\vec N$
lies in the upper three-dimensional subspace the system is
antiferromagnetic, while if it lies in the lower subspace the system
is superconducting. If the expectation value of $\vec N$ is zero the
system is neither superconducting nor antiferromagnetic. (And a fourth
possibility, seemingly not realized in nature, is that the system
could in principle be in a state which breaks both superconductivity
and antiferromagnetism.)

Zhang's work suggested that the SO(5) symmetry
is explicitly broken by small terms, in a similar way to the explicit
breaking
of chiral symmetry by small quark masses; as a result some of the
Goldstone bosons which would arise due to the breaking of SO(5) would,
in fact, be pseudo-Goldstone bosons (low-energy but not quite
massless); low-energy excitations seen in high-temperature
superconductors were among Zhang's original motivations for introducing
this model.

\section{Solitons in the SO(5) Model}
Whenever a model exhibits spontaneous symmetry breaking, there is a
family of equivalent vacua (which are rotated into one another by the
broken symmetry generators). The possibility then arises that
topological solitons could exist. General topological arguments can be
applied to any case to see if, in fact, solitons are realized.

One of the simplest examples of solitons is superconducting
vortices. These appear in the appropriate GL model, or equivalently in
the Abelian Higgs model (\ref{gl}), in 2+1 dimensions. 
The potential is the familiar
Mexican-hat potential, with a ring of vacua given by $|\phi|=v$. For
finiteness of energy, $\phi$ must go to a vacuum at infinity, but
there is no need for this to be the same vacuum along different
directions. We can construct a configuration such that the phase of
$\phi$ changes by $2\pi$ as we go around a circle at infinity; such a
configuration cannot be unwound by continuous deformations (without
wandering away from the vacuum at infinity, which would cost an
infinite energy). Furthermore, if the field configuration is
continuous, it is a topological necessity that somewhere there must be
a zero of the field. In the simplest, most rotationally symmetric
configuration, this zero will be at the origin; we may write
\be
\phi(r,\theta)=v\,f(r)e^{i\theta},
\ee
where the function $f(r)$ interpolates from zero at the origin to 1
at infinity.

In the SO(5) theory, the order parameter is considerably more
complicated, and interesting and exotic possibilities for
solitons might arise, as we will now see.\cite{abks}

The potential is assumed to be exactly invariant under rotations of
the superconducting and antiferromagnetic order parameters, and we may
assume that $V$ has the following form, 
depending only on the magnitudes of these order parameters,
$\phi=|\phi|$ and $n=|\vec n|$:
\be
V(\phi,n)=-{m_1^2\over2}\phi^2+{\lambda_1\over4}\phi^4
-{m_2^2\over2}n^2+{\lambda_2\over4}n^4
+{\lambda_3\over2}\phi^2 n^2 +\ const.
\label{potl}
\ee
Here, we have added a constant so that the minimum of the potential is
zero, included even terms up to fourth order, and
assumed the quadratic terms are such that symmetry breaking in both
sectors is favoured. The potential is assumed bounded below at all
directions at infinity, which will be true if the quartic couplings
satisfy $\lambda_{1,2}>0$ and $\lambda_3>-\sqrt{\lambda_1\lambda_2}$.

Suppose furthermore that the parameters of the potential are such that
$V$ is as shown in Fig. 1. There are two important features. First,
the global
minimum is at a nonzero value of $\phi$ with $n=0$; this corresponds
to having a superconducting ground state. Second, if we were to
force $\phi$ to be zero, the potential $V(0,n)$ is minimized at a
nonzero value of $n$. These features do occur if the parameters obey
the following conditions:
\be
{\lambda_3\over\lambda_1}>{{m_2}^2\over{m_1}^2},\qquad
{{m_1}^4\over\lambda_1}>{{m_2}^4\over\lambda_2}.
\label{conditions}
\ee
The second feature is no mere mathematical
curiosity, since at the core of a superconducting vortex $\phi$ is
indeed zero. Thus, if the potential energy had its way, the
superconducting vortex core would surely be antiferromagnetic.
%%%%%%%%%%%%%%%%%%%%%%%%%%%%%%%%%%%%%%
%%%  figure 1
%%%%%%%%%%%%%%%%%%%%%%%%%%%%%%%%%%%%%%
\begin{figure}
\epsfysize = 3in
\centerline{\epsfbox{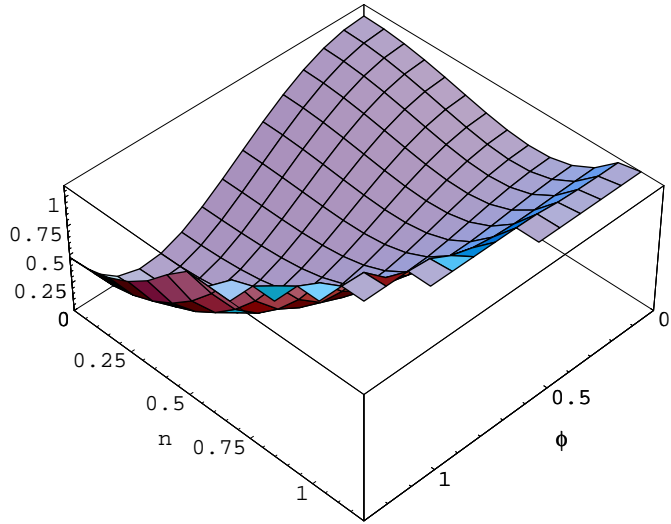}}
\baselineskip=16pt
\caption{SO(5) potential $V(\phi,n)$.}
\end{figure}
In fact, the energetics is somewhat more complicated, and there is a
range of parameters where the core is antiferromagnetic; outside this
range the core is normal.

To see this, we must solve the coupled equations for $\phi$ and
$n$. These equations come from the energy functional (we assume planar
geometry):
\bea
E&=\int\,d^{\,2}x&\left\{
{|\nabla\phi|^2\over2}-{{m_1}^2|\phi|^2\over2}
+{\lambda_1|\phi|^4\over4}\nonumber\right.\\
&&+\left.{|\nabla\vec n|^2\over2}-{{m_2}^2|{\vec n}|^2\over2}
+{\lambda_2{\vec n}^4\over4}
+{\lambda_3|\phi|^2|{\vec n}|^2\over2}
\right\}
\label{energy}
\eea

The equations of motion are straightforward, and can be rewritten in
the following form with a rotationally symmetric ansatz
$\phi(x)=v \phi(r)e^{i\theta}$, $\vec n(x)=\vec n_0 n(r)$ (where $\vec
n_0$ is a constant unit vector),
and with appropriate field and coordinate
rescalings:
\bea
{d^2\phi\over du^2}+{1\over u}{d\phi\over du}
+\left(1-{1\over u^2}\right)\phi-\delta n^2\phi-\phi^3&=&0,
\label{phieqn}\\
{d^2n\over du^2}+{1\over u}{dn\over du}
+\beta n-\alpha  \phi^2 n-n^3&=&0,
\label{neqn}
\eea
where the constants $\alpha,\ \beta$ and $\delta$ are
$\alpha=\lambda_3/\lambda_1$,
$\beta={m_2}^2/{m_1}^2$ and $\delta=\lambda_3/\lambda_1$.

The equations can be solved subject to four boundary conditions:
$\phi$ must go to zero at the origin and to 1 at infinity, and $n$
must have zero slope at the origin and must go to zero at infinity.
The crucial observation, however, is that there is no need for $n$ to
be zero at the origin, and we find that for some parameters $n$ goes
to a nonzero value, corresponding to an antiferromagnetic core for the
vortex.

%%%%%%%%%%%%%%%%%%%%%%%%%%%%%%%%%%%%%%
%%%  figure 2
%%%%%%%%%%%%%%%%%%%%%%%%%%%%%%%%%%%%%%
\begin{figure}
\epsfysize = 2.4in
\centerline{\epsfbox{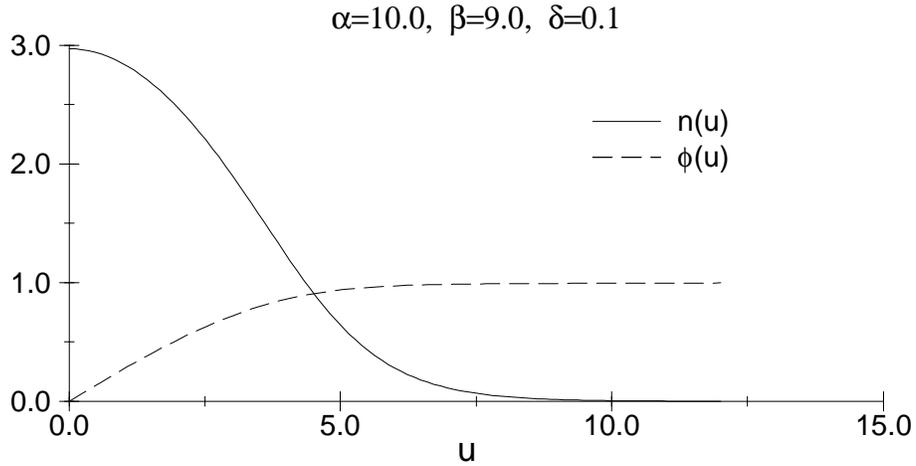}}
\baselineskip=16pt
\caption{Soliton profile.}
\end{figure}

As an example, Figure 2 displays the functions $\phi(u)$ and $n(u)$
($u$ being a scaled radial variable), for specific values of the
parameters. Since $n$ is nonzero at $u=0$, the core of the vortex in
this case is superconducting. For other values of the parameters (for
example, if $\beta$ is reduced to a sufficiently low value), one finds
that $n(u)=0$ for all $u$, indicating a normal core.

Still to be done is a systematic scan of parameter space to see under
what conditions vortices do indeed have antiferromagnetic cores. It
is also important to make contact between the parameters of the GL
model above and experimental parameters of exotic superconductors.

Finally, it is not difficult to see that other types of solitons could
in principle have similar exotic structure.\cite{climac}
One example would occur in
the antiferromagnetic phase of these materials, if the
antiferromagnetism is of an ``easy-plane'' variety. (This means that,
rather than being truly isotropic, a plane is favoured for
antiferromagnetism, due to crystal asymmetry.) In this case, it is
easy to imagine antiferromagnetic vortices of a very similar structure
to the superconducting ones discussed above, and depending on the
parameters it is possible that such vortices would have
superconducting cores.

A second example might actually be observed in certain underdoped
high-temperature superconductors which display
striping. In these materials, the striping can be understood in terms
of the formation of antiferromagnetic domain walls (domain boundaries,
really, since the materials are effectively planar), with
superconductivity occuring in the domain wall. This might fit in
nicely with the SO(5) model, since it is fairly straightforward to
construct antiferromagnetic domain boundaries which have a
superconducting core.

\end{document}